Research Paper

# Photometry of comet 29P/Schwassmann–Wachmann 1 in 2012–2019

Olena Shubina [a,b,*], Valery Kleshchonok [c], Oleksandra Ivanova [a,b,c], Igor Luk'yanyk [c], Alexander Baransky [c]

[a] *Astronomical Institute of the Slovak Academy of Sciences, 05960 Tatranská Lomnica, Slovak Republic*
[b] *Main astronomical observatory of National academy of sciences, 27 Akademika Zabolotnoho St., Kyiv, 03143, Ukraine*
[c] *Astronomical Observatory of Taras Shevchenko National University of Kyiv, 3 Observatorna St., Kyiv, 04053, Ukraine*

## ABSTRACT

The analysis of photometrical observations of comet 29P/Schwassmann–Wachmann 1 is presented. The broadband observations were carried out for 15 nights from 2012 to 2019 at the Lisnyky observational station of Taras Shevchenko National University of Kyiv. Apparent magnitudes and dust productivity level $Af\rho$ in filter R were calculated. Middle and height dust activity of the comet is characterized by parameter $Af\rho$ which varied from 1246 to 17563 cm during all periods of observation. Based on the morphological analysis, four jet-like structures were detected in the coma on almost all dates. Using the geometrical model for the jet structure interpretation during all observation sets, we obtained following results: the nucleus rotation period of $57 \pm 2$ days, the rotational axis orientation, the locations of active regions for four jet-like structures within a narrow belt near the equator.

## 1. Introduction

Comet 29P/Schwassmann–Wachmann 1 (hereafter, 29P) was discovered on 1927 November 15 by astronomers Arnold Schwassmann and Arno Arthur Wachmann (Berman and Whipple, 1928). It was the first object belonging to the small bodies of the Solar system, which was observed outside the orbit of Jupiter. It was noted that the coma had never completely disappeared in spite of differences in cometary activity levels during the whole period of observations (Jewitt, 1990). This almost constant activity has led to 29P being classified as a comet. However, the further discovery of other bodies beyond the orbit of Jupiter led to the discovery of a new population of so-called Centaurs. Centaurs are icy bodies with orbits between Jupiter and Neptune, which dynamically link the Jupiter-family comets (JFC) to the trans-Neptunian objects of the outer Solar system. Accordingly, Comet 29P was assigned to Centaur population (Jewitt, 2009). An almost circular orbit, its small inclination and the Tisserand parameter ($T_J = 2.984$ from JPL website[1]) indicate a unique feature of 29P — duality, i.e. lies between two populations of objects (Fernández et al., 2018). The orbital parameters of 29P according to JPL K192/61 are the following: a semi-major axis is 6.0002 au, eccentricity is 0.0436, the orbital period is 14.698 years, orbit inclination is 9.38°. The last perihelion the comet suffered on March 7, 2019.

The duality of the comet is also indicated by its physical properties, found from ground-based observations. It is known that for cometary nuclei the albedo value usually varies from 0.02 to 0.12, the average value is 0.07 (Jewitt, 1991). But the geometric albedo of 29P, found, for example, from photometry in the visible spectral range, is $pV = 0.13$ (Cruikshank and Brown, 1983). In Table 1 of Miles (2016) other measurements up to 0.17 are given. The value 0.33 is given on Jet Propulsion Laboratory. These are completely atypical values for a cometary nucleus. At the same time, this albedo value is characteristic of objects of the centaur group (Barucci et al., 2004). Nevertheless, 29P constantly exhibits a cometary nature, regularly outbursting. The study of morphology using digital filters shows the existence of developed structures in the coma. The presence of such structures makes it possible to determine the period of rotation of the nucleus of comet 29P. Table 1 of Miles (2016) shows the values of the rotation period determined by 4 methods: (i) inner coma photometry during quiescence, (ii) jet morphology, (iii) outburst times, and (iv) dust production models. However, Miles, based on a wide range of estimates, points out that, the rotational characteristics of the nucleus have proven to be rather elusive. He himself, proceeding from the consideration that outbursts are triggered by solar heating, the analysis yields a value for the mean solar day of $57.71 \pm 0.06$ d, equivalent to a sidereal rotation period of $57.09 \pm 0.06$ d.

Comet 29P was observed during several observing sessions in 2012–2019. The main purposes were to analyze the level of dust productivity

---

* Corresponding author at: Astronomical Institute of the Slovak Academy of Sciences, 05960 Tatranská Lomnica, Slovak Republic.
  *E-mail address:* oshubina@ta3.sk (O. Shubina).
[1] https://ssd.jpl.nasa.gov/tools/sbdb_lookup.htmlx#/?sstr=29P

O. Shubina et al.

**Table 1**
Observation log of comet 29P/Schwassmann–Wachmann 1. The observation date, total exposure time (Exp. time), heliocentric ($r$) and geocentric ($\Delta$) distance, phase angle ($\alpha$), the position angles of the extended Sun-to-target radius vector ($\phi$).

| Year | Date | Exp. time, s | $r$, au | $\Delta$, au | $\alpha$, deg | $\phi$, deg |
|---|---|---|---|---|---|---|
| 2012 | Jan 30 | 1560 | 6.261 | 5.786 | 8.22 | 297.26 |
|  | Mar 07 | 1680 | 6.260 | 5.362 | 4.18 | 312.44 |
| 2013 | Apr 17 | 1320 | 6.222 | 5.240 | 2.30 | 343.56 |
|  | Jun 20 | 1560 | 6.211 | 5.645 | 8.16 | 105.78 |
| 2018 | Aug 17 | 1500 | 5.775 | 4.844 | 4.32 | 234.00 |
|  | Aug 26 | 570 | 5.774 | 4.797 | 2.81 | 225.05 |
|  | Aug 30 | 1080 | 5.774 | 4.784 | 2.16 | 217.03 |
|  | Sep 14 | 840 | 5.773 | 4.776 | 1.46 | 116.51 |
|  | Oct 09 | 1200 | 5.771 | 4.910 | 5.44 | 77.72 |
|  | Oct 12 | 900 | 5.771 | 4.937 | 5.88 | 76.62 |
|  | Oct 13 | 1140 | 5.771 | 4.946 | 6.02 | 76.82 |
| 2019 | Aug 22 | 1440 | 5.772 | 5.084 | 7.85 | 239.60 |
|  | Oct 01 | 1200 | 5.775 | 4.795 | 2.24 | 196.37 |
|  | Oct 18 | 1200 | 5.776 | 4.807 | 2.49 | 109.79 |
|  | Oct 30 | 1140 | 5.778 | 4.867 | 4.27 | 88.02 |

**Table 2**
Apparent magnitudes of comet 29P/Schwassmann–Wachmann 1 obtained within different apertures.

| Date | Average aperture size, $\times 10^3$ km | | | |
|---|---|---|---|---|
|  | 21 | 40 | 54 | 67 |
| 2012 | | | | |
| Jan 30 | 17.82 ± 0.19 | 16.81 ± 0.15 | 16.32 ± 0.12 | 16.05 ± 0.16 |
| Mar 07 | 17.55 ± 0.11 | 17.42 ± 0.11 | 17.38 ± 0.12 | – |
| 2013 | | | | |
| Apr 17 | 16.89 ± 0.09 | 15.50 ± 0.09 | 15.02 ± 0.09 | 14.70 ± 0.09 |
| Jun 20 | 15.84 ± 0.10 | 14.72 ± 0.11 | 14.09 ± 0.08 | 13.71 ± 0.06 |
| 2018 | | | | |
| Aug 17 | 16.59 ± 0.11 | 16.27 ± 0.11 | 16.04 ± 0.12 | 15.89 ± 0.11 |
| Aug 26 | 16.36 ± 0.08 | 16.12 ± 0.11 | 16.01 ± 0.12 | 15.88 ± 0.11 |
| Aug 30 | 16.32 ± 0.06 | 16.03 ± 0.07 | 15.87 ± 0.12 | 15.64 ± 0.13 |
| Sep 14 | 15.97 ± 0.05 | 15.61 ± 0.05 | 15.44 ± 0.06 | 15.33 ± 0.04 |
| Oct 09 | 15.41 ± 0.01 | 14.89 ± 0.02 | 14.54 ± 0.02 | 14.23 ± 0.03 |
| Oct 12 | 16.02 ± 0.04 | 15.42 ± 0.04 | 15.10 ± 0.04 | 14.86 ± 0.04 |
| Oct 13 | 16.41 ± 0.05 | 15.78 ± 0.05 | 15.38 ± 0.05 | 15.08 ± 0.06 |
| 2019 | | | | |
| Aug 22 | 16.96 ± 0.10 | 16.54 ± 0.10 | 16.33 ± 0.09 | 16.11 ± 0.10 |
| Oct 01 | 16.49 ± 0.03 | 16.01 ± 0.02 | 15.80 ± 0.04 | 15.63 ± 0.03 |
| Oct 18 | 16.11 ± 0.05 | 15.24 ± 0.03 | 15.01 ± 0.04 | 14.86 ± 0.02 |
| Oct 30 | 16.63 ± 0.13 | 16.01 ± 0.24 | 15.69 ± 0.32 | 15.48 ± 0.36 |

and compare it with the same for other active objects of the Solar system, as well as, based on morphological analysis and modeling, to determine some parameters of the comet's nucleus and active regions on its surface. So Section 2 is devoted to the circumstances of observation. Section 3 is devoted to computing the $Af\rho$ parameter and its analysis from a series of observations in different periods of the comet's activity, the morphological analysis of photometric images and modeling of its jet structure.

## 2. Observations and data processing

The photometry observations of comet 29P were carried out at the observation station of Taras Shevchenko National University of Kyiv located in Lisnyky (the international code is 585). The data were obtained using the telescope AZT-8 ($D = 70$ cm, $F = 2.8$ m) in 2012, 2013, and during 2018–2019. CCD FLI PL4710 was used as a detector. The chip size is $1027 \times 1056$ pixels. The pixel scale was $0.95''$/pixel. Exposure times ranged from 30 to 120 s and were taken using a broadband Johnson-Cousins R filter. During observations, the weather conditions changed. In our analysis we used only the data with seeing values of 3–4$''$ More details on the comet observation circumstances are collected in Table 1.

We processed the obtained data using standard procedures: bias and dark subtraction, flat-field correction. Also, the images were cleaned from cosmic particle marks. The subtraction of the sky level was made by using the standard IDL procedure. In addition, the entire series of homogeneous data per one observation night were summed for further use. For the photometric calibration of our data, we used the stars over the field of view. The stellar magnitudes of the standard stars were taken from the catalog UCAC4 (Zacharias et al., 2013), which accuracy is 0.05–0.1 mag.

The apparent magnitude is determined as follows:

$$m_c(\lambda) = -2.5 \cdot \lg \left[ \frac{I_c(\lambda)}{I_s(\lambda)} \right] + m_s - 2.5 \cdot \lg P(\lambda) \cdot \Delta M, \quad (1)$$

where $m_c$ and $m_s$ is the apparent magnitude of the comet and the star, respectively; $I_c(\lambda)$ and $I_s(\lambda)$ is intensity in relative units for the comet and the star, respectively; $P$ is sky transparency, $\Delta M$ is the airmass difference between the comet and the star. As far as we used stars on the same frame as the comet we took $\Delta M$ equal to zero.

To estimate a dust productivity level we used the $Af\rho$ value (A'Hearn et al., 1984), where $A$ is the average grain geometric albedo, $f$ the filling factor in the aperture field of view and $\rho$ the linear radius of the aperture at the comet. It can be derived from the calculated photometric dust coma magnitude obtained previously. According to Mazzotta Epifani et al. (2010), we calculated $Af\rho$ with:

$$Af\rho = \frac{4r^2\Delta^2}{\rho} \cdot 10^{0.4(m_{Sun}-m_c)}, \quad (2)$$

where $m_c$ and $m_{Sun}$ is the comet magnitude and the solar magnitude in the corresponding band, $r$ is the heliocentric distance in astronomical units, $\Delta$ is the geocentric distance expressed in cm and $\rho$ is the aperture radius in cm. $Af\rho$ is often used to compare the dust production rate between comets, but it should be noted that it uses observational and physical parameters that could change in different cases. Nevertheless, as a general rule, it is assumed that large $Af\rho$ values indicate high dust activity. We observed comet 29P within broadband R filter, which might include gas emission species. Spectral investigations of the comet in the optical wavelength region revealed the presence of weak emissions of CO, CN, and $N_2^+$ molecules (see Ivanova et al., 2018 and reference therein). Nevertheless, Picazzio et al. (2019) reported that gas contribution is about ∼1% within V filter. So we accepted that the gas contribution is negligible in R band. Taking this into account our estimation of dust productivity with $Af\rho$ proxy is appropriate and can be used to study a dust component of the cometary coma.

## 3. Results and analysis

### 3.1. Photometry

Using Eq. (1) we calculate apparent magnitudes within four diaphragms. The results of our calculation within different apertures are collected in Table 2.

Based on calculated magnitudes we computed the $Af\rho$ parameter (Eq. (2)) using the same aperture sizes and the Sun magnitude value in R filter of $-27.26$. The $Af\rho$ parameter is dependent on the phase angle of the comet, thus we applied the phase correction proposed by Schleicher et al. (1998) for Comet Halley. The dust production measurement corrected for the phase angle is denoted as $A(0°)f\rho$. It is the phase-corrected dust production results that are presented in Table 3.

As one can see from Table 3, the parameter value changes with the aperture sizes. We can specify three types based on the trend behavior of $Af\rho$ parameter: increasing the parameter value with the aperture increases, decreasing the parameter value, and a mixed case.

**Table 3**

$A(0°)f\rho$ parameter in cm for comet 29P/Schwassmann–Wachmann 1 obtained within corresponding apertures.

| Date | Average aperture size, ×10³, km | | | |
|---|---|---|---|---|
| | 21 | 40 | 54 | 67 |
| 2012 | | | | |
| Jan 30 | 1246 ± 219 | 1895 ± 259 | 2127 ± 237 | 2120 ± 320 |
| Mar 07 | 1268 ± 128 | 858 ± 120 | 636 ± 64 | – |
| 2013 | | | | |
| Apr 17 | 2088 ± 177 | 4499 ± 383 | 4996 ± 425 | 5219 ± 444 |
| Jun 20 | 7405 ± 652 | 12464 ± 1295 | 15877 ± 1198 | 17563 ± 982 |
| 2018 | | | | |
| Aug 17 | 1774 ± 174 | 1592 ± 163 | 1477 ± 158 | 1350 ± 142 |
| Aug 26 | 2090 ± 152 | 1739 ± 183 | 1438 ± 165 | 1303 ± 135 |
| Aug 30 | 2070 ± 113 | 1814 ± 120 | 1580 ± 179 | 1561 ± 190 |
| Sep 14 | 2744 ± 120 | 2551 ± 113 | 2233 ± 115 | 1986 ± 76 |
| Oct 09 | 5545 ± 51 | 5983 ± 87 | 6191 ± 112 | 6546 ± 182 |
| Oct 12 | 3302 ± 126 | 3830 ± 134 | 3845 ± 139 | 3829 ± 150 |
| Oct 13 | 2299 ± 98 | 2752 ± 133 | 2967 ± 147 | 3123 ± 168 |
| 2019 | | | | |
| Aug 22 | 1535 ± 145 | 1505 ± 138 | 1365 ± 116 | 1339 ± 128 |
| Oct 01 | 1784 ± 53 | 1846 ± 41 | 1683 ± 63 | 1572 ± 46 |
| Oct 18 | 2537 ± 123 | 3770 ± 104 | 3478 ± 127 | 3213 ± 72 |
| Oct 30 | 1715 ± 203 | 2028 ± 450 | 2040 ± 596 | 1992 ± 664 |

The first group includes the results of 2012-01-30, 2013-04-17, 2013-06-20, 2018-10-09, 2018-10-13. The common thing for all these data is the presence of at least a significant extended coma. Moreover, it was reported about the outburst on the comet on 2013-06-20 (Keane et al., 2013). There are no published sources with information about outbursts on other dates. But based on our investigations we may conclude that 29P experienced a high increase in the brightness in September–October 2018. As for the rest two dates (2013-01-30 and 2013-04-17), based on images from the website of MISAO project (Seiichi Yoshida's home page[2]) one can conclude that during these periods at least the increase in activity happened or even some outbursts. The second group includes results obtained on 2012-03-07, August–September in 2018, and 2019-08-22. On these dates, the coma was very weak (according to images from the MISAO project website). But it should be noted that from 2018-09-27, the comet's brightness began to increase, and the coma started to expand. The third group does not show a uniform trend, i.e. the dust activity level demonstrates the increase within smaller apertures, but within bigger ones, the trend changes to decreasing. We detected such behavior of the $Af\rho$ parameter in October 2019, when the coma was well-marked but not so significant. Our assumptions about the increase of the cometary activity of 29P during our observations are pretty good correlated with the data published on the website of the British Astronomical Association.[3] Based on reduced light curves, one can see the dramatic change in the brightness in September–October 2018. It should be noted that during the brightness decrease after the outburst, there were a few smaller outbursts, that can explain the observed activity. Based on the light curves of comet 29P mentioned above, one can see that most of our observations were carried out on the dates when the comet did not suffer significant outburst activity. The exception is the data obtained in September–October 2018.

Our results of the dust productivity level are also in good agreement with the published ones. It can be seen from Fig. 1, where we compiled literature dates from 2007 to 2020 with our measurements computed within the aperture radius of about 21 000 km. To compare the results we used $Af\rho$ values obtained for the observed phase angle. Only in 2013, we did not find any refereed results to compare. It also should be noted that these values mostly were obtained in the R filter. But we included some data points obtained in the V filter to complete the data set (e.g. the result in 2015 Garcia et al., 2020). From this plot, we clearly see some general trends in dust productivity levels. The highest value was detected in 2008 after a major outburst as the author mentioned (Schambeau et al., 2017). It may be assumed that during 2013–2015 the local maximum should also be presented. Furthermore, the increase in dust productivity level in 2018 with the following decrease in 2019 is remarkable. All of this analysis confirms the challenging nature of comet 29P.

Fig. 2 demonstrates the comparison of $Af\rho$ parameter obtained in this paper for comet 29P with the values obtained within R filters for small bodies belonging to different groups. We selected about 20 JFC observed beyond 3 au (Snodgrass et al., 2008; Lowry and Fitzsimmons, 2001, 2005; Lowry and Weissman, 2003; Lowry et al., 2003, 1999; Lamy et al., 2009; Mazzotta Epifani et al., 2007; Snodgrass et al., 2006), 18 active centaurs (Rousselot et al., 2016, 2021; Wong et al., 2019; Jewitt, 2005), 8 active asteroids (Martino et al., 2019; Meech et al., 2009; Neslusan et al., 2016; Ferrín et al., 2012; Hsieh and Sheppard, 2015; Hsieh et al., 2013, 2012; Shi et al., 2019; Kokhirova et al., 2018), and 11 long-periodic comets (A'Hearn et al., 1995; Schleicher et al., 1997; Szabó et al., 2001; Mazzotta Epifani et al., 2009; Korsun et al., 2014, 2016; Hui et al., 2018). If the authors obtained $Af\rho$ parameters in several apertures, we chose the one that size is the closest to our diaphragm in kilometers. We should note that we consider results obtained only at heliocentric distances larger than 3 au due to the strong effect of the gas component of comae at closer distances. The exception is active asteroids, where the gas contribution is absent or weak. As we can see in this plot, long-periodic comets tend to have higher $Af\rho$ values, while JFC demonstrate lower ones. Active asteroids concentrate at 2–4 au and show $Af\rho$ values mainly around hundred centimeters, except two points corresponded to the asteroid (596) Scheila (Neslusan et al., 2016; Kokhirova et al., 2018), that demonstrated dust productivity levels about 2900 and 1400 cm. Meanwhile, centaurs demonstrate a wide dispersion of heliocentric distances, but their $Af\rho$ are within a region of 100–1000 cm. At large distances from the Sun, it seems to be some plateau. It might be caused due to lack of data, and future observations and investigations of active centaurs at large heliocentric distances shed light on this issue. As one can see from Fig. 2 our measurements of dust productivity level are higher than ones for JFC and tend to be active centaurs. According to Fig. 1 our results can be considered as the mean value for comet 29P. Thus, if we analyze only dust productivity using the $Af\rho$ proxy, then it is challenging to designate 29P as a centaur or a long-period comet. In this case, the object reveals its duality. But if one takes the dynamical properties of this comet into account, then it is seen that 29P tends to be classified as a centaur.

### 3.2. Morphology

We constructed intensity maps by averaging all the photometric images obtained on different nights separately. For each date, the coma demonstrated asymmetry, especially during brightness increases. In order to reveal the low-contrast structures in the dust coma, the images were treated with a digital filter. We applied $1/\rho$ profile (Samarasinha and Larson, 2014), which can be used for dusty comets. This technique enabled us to remove the bright background from the cometary coma and highlight the low-contrast features. To double-check we also used azimuthal average and renormalization methods (Samarasinha and Larson, 2014), but their results were the same as the $1/\rho$ profile showed. The usage of a rotational gradient method (Larson and Sekanina, 1984) confirmed existence of the jet-like structures in the cometary coma. Fig. 3 demonstrates results for eight observed nights with the best photometrical accuracy. The comet demonstrated an extended coma on 2013-06-20 and 2018-10-13. In the former case, it was reported about the outburst (Keane et al., 2013). In the latter case, we definitely see an increase in activity and brightness. Our observations in October 2018 were obtained during brightness decreasing, right after the repeated outburst (see the light curves of comet 29P published on the website of the British Astronomical Association mentioned in Section 3). Although on the other dates the comet did not demonstrate prominent activity, nevertheless, small structures can be identified: one on 2018-09-14, several on 2012-01-30, and weak activity on 2019-10-30.

---

[2] http://www.aerith.net/
[3] https://britastro.org/section_information_/comet-section-overview/mission-29p-centaur-comet-observing-campaign

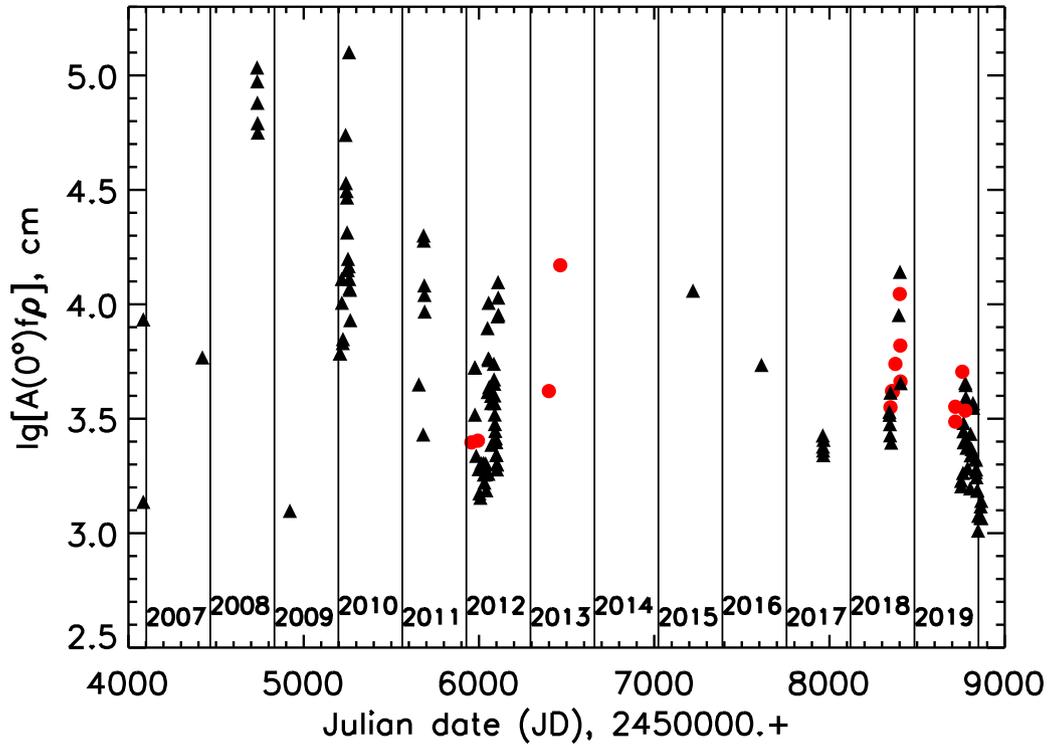

**Fig. 1.** Compilation of A(0°)fρ parameter of comet 29P/Schwassmann–Wachmann 1 obtained mostly in *R* filter based on results of Clements and Fernandez (2021), Garcia et al. (2020), Hosek et al. (2013), Ivanova et al. (2009, 2011, 2016), Kokhirova et al. (2020), Miles et al. (2016), Picazzio et al. (2019), Schambeau et al. (2017), Shi et al. (2014), Trigo-Rodríguez et al. (2010, 2012), Voitko et al. (2022) (black triangles), and this work (red dots). Red dots represent our results within the aperture of about 21 000 km.

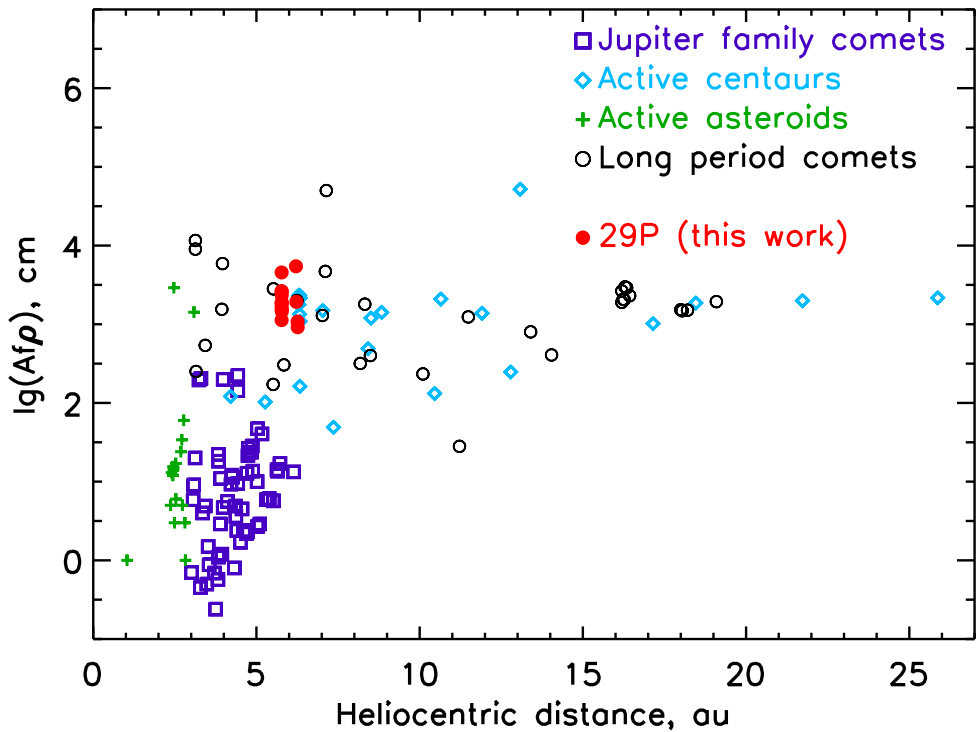

**Fig. 2.** Afρ parameter (obtained in *R* filter) of comet 29P/Schwassmann–Wachmann 1 (this work) and different dynamical groups depending on the heliocentric distance. Dust activity of different groups shows by corresponding symbols indicated in the figure. Data for JFC, long-period comets, active asteroids and centaurs are taken from Snodgrass et al. (2008), Lowry and Fitzsimmons (2001, 2005), Lowry and Weissman (2003), Lowry et al. (2003, 1999), Lamy et al. (2009), Mazzotta Epifani et al. (2007), Snodgrass et al. (2006), Rousselot et al. (2016, 2021), Wong et al. (2019), Jewitt (2005), A'Hearn et al. (1995), Schleicher et al. (1997), Szabó et al. (2001), Mazzotta Epifani et al. (2009), Korsun et al. (2014, 2016), Hui et al. (2018), Martino et al. (2019), Meech et al. (2009), Neslusan et al. (2016), Ferrín et al. (2012), Hsieh and Sheppard (2015), Hsieh et al. (2013, 2012), Shi et al. (2019), Kokhirova et al. (2018).

### 3.3. Modeling of coma jet structure

A rotating period of a nucleus is an important parameter that must be taken into account in the modeling of a jet structure. One can determine this parameter using photometrical data or interpretation of structural components of a coma. It is a significant difference between rotating periods of comet 29P determined in literature. Earlier estimations of the period were based on photometrical results and revealed dramatically smaller values, e.g., $120 \pm 1$ h (Whipple, 1980), 6 days (Jewitt, 1990), $10 \pm 1$ h (Luu and Jewitt, 1993), and 14 hours (Meech et al., 1993). More recent investigations of coma structure variations showed longer rotation periods of a nucleus. For instance, Stansberry et al. (2004) estimated the rotation period of comet 29P as 60 days, Trigo-Rodríguez et al. (2010) estimated the period of 50 days. Ivanova et al. (2012) obtained two values for the period $12.1 \pm 1.2$ and $11.7 \pm 1.5$ days for the observations taken in December 2008 and February 2009, respectively. The authors also determined the inclination of the rotation axis to the comet's orbit plane as 65°. Based on the analysis of 64 outbursts over the period 2002–2014 (Miles, 2016) found an activity's periodicity of $57.6 \pm 0.4$ d. The presence of long-lived active sources of outburst is noted at the same time. These sources are localized in the range of longitudes 135–150°. Miles (2016) also noted that the jet structure is disrupted during outbursts, which can lead to incorrect conclusions when interpreting it. The images from the Hubble Space Telescope in 1996 (Feldman et al., 1996) showed 6 active regions that generated outbursts. The existence of several active regions in combination with outburst activity leads to such dispersion of the value of the comet's nucleus rotation period in different authors.

A cometary coma does not have a spherical form. Moreover, it changes over time. It is often observed that the coma demonstrates oblongness in the sun-ward direction or the perpendicular one. Such behavior in the coma indicates the presence of a large jet structure. But obtained images do not reveal separated jets. It might be caused by insignificant differences between locations of the active regions on the nucleus. Thus, one can observe the jets' overlapping due to the certain jet width and the velocity dispersion. This issue adds challenges to the jet interpretation. The comparison of coma features obtained on different dates can help. But for this, one should assume the long-term existence of active regions which produce these jets. And such behavior of active regions is a point indeed for comets. For example, the existence of long-term active regions was shown for short-period comet 2P/Encke (Sekanina, 1988b,a). Also, the disk-shaped form of the coma was explained by the jet activity during the whole observable period for hyperbolic comet C/2014 B1 (Schwartz) (Jewitt et al., 2019) and distant comet C/2011 P1 (McNaught) (Korsun et al., 2016).

To model jets, we used a simple geometrical model, which has restrictions, but was successfully applied to analyze structures in a coma of comets 2P/Encke (Rosenbush et al., 2020), C/2011 KP36 (Spacewatch) (Ivanova et al., 2021), and 46P/Wirtanen (Rosenbush et al., 2021). This model shows only the position of the jet's axis at each time moment. The model takes into account mainly geometric factors: positions of the comet, the Earth, and the Sun, the orientation of the nucleus rotation axis, and the location of active areas on the nucleus. The active area size, the size distribution of particles, and the acceleration of dust particles under the action of solar radiation are not considered. This leads to a limitation on the model jet length. The simulation is carried out up to a cometocentric distance where the acceleration of solar radiation can be neglected, i.e., an additional offset of a particle caused by the sun radiation is small compared to the cometocentric distance. The model assumes the termination of the matter outflow from the active area after the termination of insolation. Important parameters of the geometric model are the nucleus rotation period and the matter velocity in the jet. They determine the symmetry axis shape of the observed jets with other geometric factors and operate as a characteristic length equal to the product of the particle velocity and the rotation period for a single observation. It is possible to determine the nucleus rotation period and the matter velocity in the jet separately for several observations on different dates, assuming that the particle velocity does not change. The doubtless advantage of this model is its simplicity. Although it does not consider many physical parameters, the model allows one to determine some of the necessary parameters of both the nucleus and the outflowing matter to reproduce the observed features of a cometary coma. The parameters obtained using the simulation can be the basis for further modeling of microphysical and macrophysical properties of the nucleus, its surface, and the cometary dust. In this sense, it is also important to use the model for simulating the jet structures of comets based on both a single comet observation with a well-developed coma and long-term observations. Thus, one can verify and improve the model, for example, considering the distribution of particles leaving the nucleus over their sizes, directions of the initial velocity vector, etc.

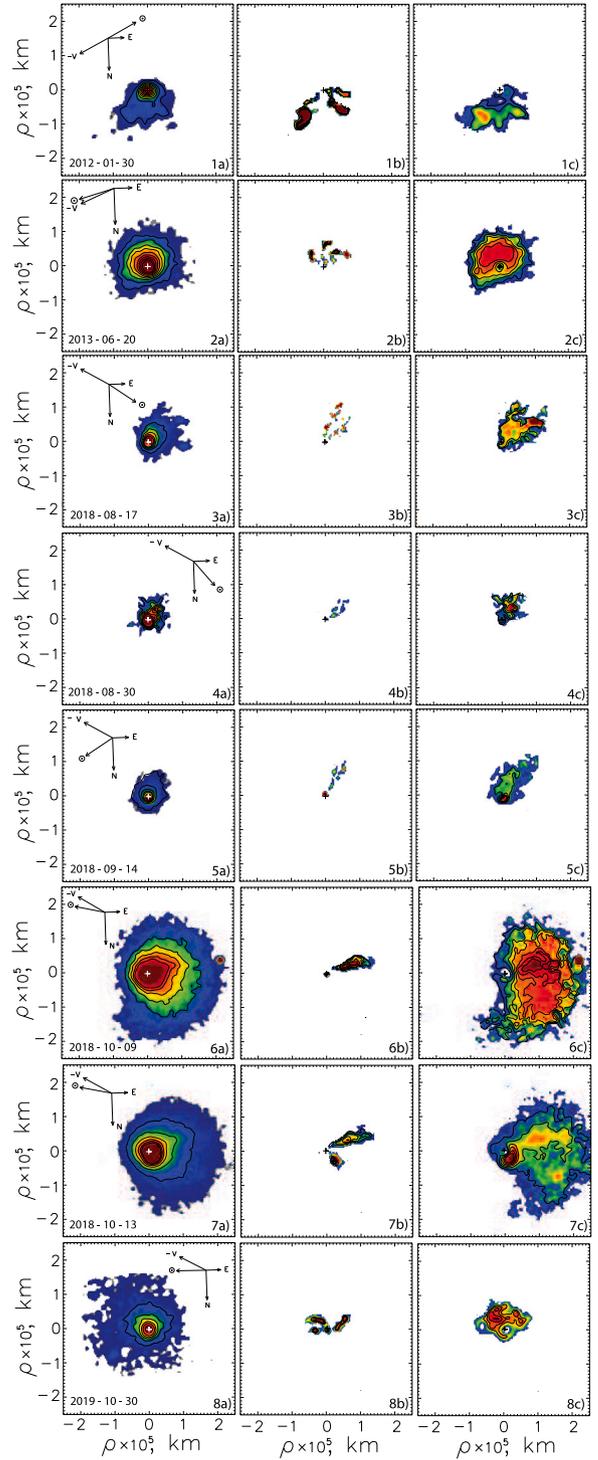

**Fig. 3.** Intensity maps of comet 29P/Schwassmann–Wachmann 1 in the $R$ filter obtained on 2012-01-30 (1), 2013-06-20 (2), 2018-08-17 (3), 2018-08-30 (4), 2018-09-14 (5), 2018-10-09 (6), 2018-10-13 (7), and 2019-10-30 (8). Panels $a$ present direct images of the comet with the isophots, $b$ — images processed by a rotational gradient method (Larson and Sekanina, 1984), and $c$ — images to which a division by $1/\rho$ profile (Samarasinha and Larson, 2014) was applied. Also, the directions to the North, the East, the Sun, and the negative heliocentric velocity vector of the comet as seen in the observer's plane of the sky are noted for each date. The optocenter indicated by the cross mark.

As shown in Section 3.2, the comet demonstrated nonuniform morphological structure. Applying our model, we simulated positions of four jets detected on processed images of comet 29P. Table 4 includes coordinates of active regions that produce observable jets. The jet marked J1 has a conventional value of the longitude of 0 degrees. The best agreement between observed images and modeled jets for all dates is when the rotation period is $57 \pm 2$ days, and the outflow velocity of matter in jets is $0.34 \pm 0.02$ km/s. Based on observations of the jet structure (Stansberry et al., 2004) determined the velocity of matter in jets to be $0.5$ km/s. The existence of jets at such large heliocentric distances was explained by the outflow of CO molecules. Such an estimate seems doubtful if we take into account the maximum velocity of CO molecules as $0.51$ km/s (Gunnarsson et al., 2008), because the dust particles velocity is lower than CO one. According to the (Stansberry et al., 2004) observations, the redefinition of the velocity made by Miles (2016) gives a slightly lower value of $0.45$ km/s, which is also quite large. The coordinates of the nucleus's north pole, which are determined by the geometric model, are $RA = 185 \pm 8°$, $DEC = 12 \pm 3°$. It corresponds to the fact that the axis of rotation is directed approximately towards the observer, and the direction of rotation is direct. Since the active areas that form the jets are located in a narrow belt near the equator, the jet structures practically lie in the sky plane. Such their location leads to the fact that their appearance is practically not distorted due to projection onto the sky plane. The collocation of the active areas likely could change due to the high activity of comet 29P. Therefore, we determined the rotation period based on observations in August–September of 2018, when the comet was in a period of low activity between the outbursts. The results of our modeling are presented in Fig. 4. All active regions were model parameters for all dates. The jet J1 is absent in several images due to the non-insolation of the respective active areas. The jets in the images for 2018-10-09 and 2018-10-13 (panels f and g of Fig. 4, respectively) are slightly visible against the increased coma background, which is caused by an active process (outburst). These findings correspond to ones obtained by Miles (2016), although the method for determining the rotation period in that work is based on the analysis of comet 29P outburst activity. Our results show that the active regions are located in a relatively narrow belt with a longitude width of about $115 \pm 20°$, which is close to the size of the active belt 135–150° (Miles, 2016). Obtained value of the rotation period $57 \pm 2$ d also is in good agreements with $57.6 \pm 0.4$ d determined by Miles (2016).

The existence of the narrow belt with active regions concentration is similar to comet 67P/Churyumov-Gerasimenko, which nucleus consists of two components. Moreover, permanent jets are observed in the area, where two pieces compound (Sierks et al., 2015). Probably if the cometary nucleus consists of two components, the active areas might locate over a "neck" region.

Table 4

Parameters of active regions created by observed jets on comet 29P/Schwassmann–Wachmann 1.

| Jet mark | Longitude, deg | Latitude, deg | Date of jet observation |
|---|---|---|---|
| J1 | $0 \pm 9$ | $-5 \pm 4$ | 2012-01-30, 2013-06-20, 2018-08-30 |
| J2 | $45 \pm 9$ | $5 \pm 4$ | all |
| J3 | $90 \pm 9$ | $6 \pm 4$ | all |
| J4 | $115 \pm 9$ | $-12 \pm 4$ | 2012-01-30, 2013-06-20, 2018: 08–17, 09–14, 10–09, 10-13 2019-10-30 |

## 4. Conclusions

In this paper, we analyzed the results of the monitoring observations of comet object 29P/Schwassmann–Wachmann 1 in the broadband filter $R$ obtained during 15 nights in 2012–2019 years. The main purposes were to analyze the level of dust productivity and compare it with the same for other active objects of the Solar system, as well as, based on morphological analysis and modeling, to determine some parameters of the comet's nucleus and active areas on its surface. The main results can be summarized as follows:

- The dust productivity, $A(0°)f\rho$, varied from $636 \pm 64$ cm within $\rho \approx 54000$ km (on 2012-03-07) when the comet was faint to $17563 \pm 982$ cm within $\rho \approx 67000$ km (on 2013-06-20) when 29P suffered the outburst. Obtained values are significantly higher than the dust productivity of Jupiter family comets but in good agreement with ones of active centaurs or long period comets. But if one takes the dynamical properties of this comet into account, then it is seen that 29P tends to be classified as a centaur based on $Af\rho$ values.
- Image analysis and the use of a digital filter showed a strong inhomogeneity of the coma and made it possible to identify 4 jet structures.
- The use of a geometric model to describe the jet structure for the entire observation period made it possible to determine some parameters of the nucleus. The rotation period of the nucleus equal $57 \pm 2$ d and the outflow velocity of the matter in the jet in $0.34 \pm 0.02$ km/s. The coordinates of the north pole of the nucleus are $RA = 185 \pm 8°$, $DEC = 12 \pm 3°$. The axis of rotation is directed approximately towards the observer with propagate direction of spin.
- Since the active areas that form the jets are located in a narrow belt near the equator with length in longitude about $115 \pm 20°$, the jet structures practically lie in the sky plane.

## CRediT authorship contribution statement

**Olena Shubina:** Software, Formal analysis, Investigation, Writing – original draft, Visualization. **Valery Kleshchonok:** Methodology, Formal analysis, Writing – review & editing. **Oleksandra Ivanova:** Conceptualization, Software, Formal analysis, Writing – review & editing, Supervision. **Igor Luk'yanyk:** Conceptualization, Writing – review & editing. **Alexander Baransky:** Carry out observations.

## Data availability

Data will be made available on request.

## Acknowledgments

This work was supported by a grant of the President of Ukraine for young scientists (OSh). The research is supported by the Slovak Academy of Sciences (grant Vega 2/0059/22) and by the Slovak Research and Development Agency under Contract no. APVV-19-0072 (OSh & OI). The research by OI, IL, VK is supported by the project 0122U001911 of the Ministry of Education and Science of Ukraine.

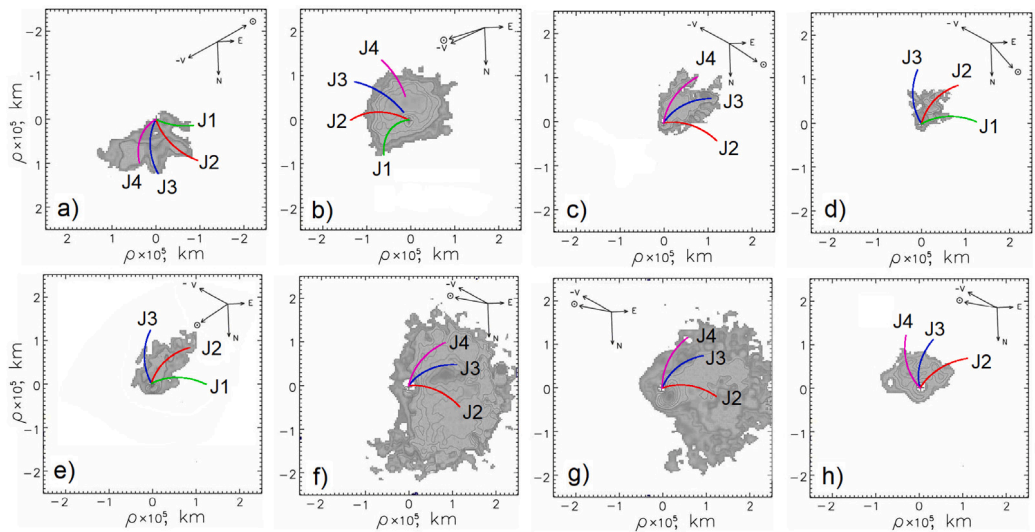

**Fig. 4.** The coma jet structures (J1, J2, J3, and J4) of comet 29P/Schwassmann–Wachmann 1 based on the geometrical model for following data 2012-01-30 (a), 2013-06-20 (b), 2018-08-17 (c), 2018-08-30 (d), 2018-09-14 (e), 2018-10-09 (f), 2018-10-13 (g), and 2019-10-30 (h). The processed images using the digital $1/\rho$ filter are used as a background for indicating jet structures.

## References

A'Hearn, M.F., Millis, R.C., Schleicher, D.O., Osip, D.J., Birch, P.V., 1995. The ensemble properties of comets: Results from narrowband photometry of 85 comets, 1976–1992.. Icarus 118 (2), 223–270. http://dx.doi.org/10.1006/icar.1995.1190.

A'Hearn, M.F., Schleicher, D.G., Millis, R.L., Feldman, P.D., Thompson, D.T., 1984. Comet bowell 1980b. Astron. J. 89, 579–591. http://dx.doi.org/10.1086/113552.

Barucci, M.A., Doressoundiram, A., Cruikshank, D.P., 2004. Surface characteristics of transneptunian objects and centaurs from photometry and spectroscopy. In: Festou, M.C., Keller, H.U., Weaver, H.A. (Eds.), Comets II. p. 647.

Berman, L., Whipple, F.L., 1928. Elements and ephemeris of comet J 1927 (Schwassmann–Wachmann). Lick Obs. Bull. 394, 117–119. http://dx.doi.org/10.5479/ADS/bib/1928LicOB.13.117B.

Clements, T.D., Fernandez, Y., 2021. Dust production from mini outbursts of comet 29P/Schwassmann–Wachmann 1. Astron. J. 161 (2), 73. http://dx.doi.org/10.3847/1538-3881/abd1d7.

Cruikshank, D.P., Brown, R.H., 1983. The nucleus of comet P/Schwassmann–Wachmann 1. Icarus 56 (3), 377–380. http://dx.doi.org/10.1016/0019-1035(83)90158-6.

Feldman, P.D., McPhate, J.B., Weaver, H.A., Tozzi, G.-P., A'Hearn, M.F., 1996. Dust outflow velocity of comet 29P/Schwassmann–Wachmann1 during outburst. In: AAS/Division for Planetary Sciences Meeting Abstracts #28. In: AAS/Division for Planetary Sciences Meeting Abstracts, vol. 28, p. 08.08.

Fernández, J.A., Helal, M., Gallardo, T., 2018. Dynamical evolution and end states of active and inactive centaurs. Planet. Space Sci. 158, 6–15. http://dx.doi.org/10.1016/j.pss.2018.05.013, arXiv:1805.05994.

Ferrín, I., Hamanowa, H., Hamanowa, H., Hernández, J., Sira, E., Sánchez, A., Zhao, H., Miles, R., 2012. The 2009 apparition of methuselah comet 107P/Wilson–Harrington: A case of comet rejuvenation? Planet. Space Sci. 70 (1), 59–72. http://dx.doi.org/10.1016/j.pss.2012.05.022, arXiv:1205.6874.

Garcia, R.S., Gil-Hutton, R., García-Migani, E., 2020. Observational results for five short-period and five long-period comets. Planet. Space Sci. 180, 104779. http://dx.doi.org/10.1016/j.pss.2019.104779.

Gunnarsson, M., Bockelée-Morvan, D., Biver, N., Crovisier, J., Rickman, H., 2008. Mapping the carbon monoxide coma of comet 29P/Schwassmann–Wachmann 1. Astron. Astrophys. 484 (2), 537–546. http://dx.doi.org/10.1051/0004-6361:20078069.

Hosek, J., Blaauw, R.C., Cooke, W.J., Suggs, R.M., 2013. Outburst dust production of comet 29P/Schwassmann–Wachmann 1. Astron. J. 145 (5), 122. http://dx.doi.org/10.1088/0004-6256/145/5/122.

Hsieh, H.H., Kaluna, H.M., Novaković, B., Yang, B., Haghighipour, N., Micheli, M., Denneau, L., Fitzsimmons, A., Jedicke, R., Kleyna, J., Vereš, P., Wainscoat, R.J., Ansdell, M., Elliott, G.T., Keane, J.V., Meech, K.J., Moskovitz, N.A., Riesen, T.E., Sheppard, S.S., Sonnett, S., Tholen, D.J., Urban, L., Kaiser, N., Chambers, K.C., Burgett, W.S., Magnier, E.A., Morgan, J.S., Price, P.A., 2013. Main-belt comet P/2012 T1 (PANSTARRS). Astrophys. J. Lett. 771 (1), L1. http://dx.doi.org/10.1088/2041-8205/771/1/L1, arXiv:1305.5558.

Hsieh, H.H., Sheppard, S.S., 2015. The reactivation of main-belt comet 324P/La Sagra (P/2010 R2). Mon. Not. RAS 454 (1), L81–L85. http://dx.doi.org/10.1093/mnrasl/slv125, arXiv:1508.07140.

Hsieh, H.H., Yang, B., Haghighipour, N., Novaković, B., Jedicke, R., Wainscoat, R.J., Denneau, L., Abe, S., Chen, W.-P., Fitzsimmons, A., Granvik, M., Grav, T., Ip, W., Kaluna, H.M., Kinoshita, D., Kleyna, J., Knight, M.M., Lacerda, P., Lisse, C.M., Maclennan, E., Meech, K.J., Micheli, M., Milani, A., Pittichová, J., Schunova, E., Tholen, D.J., Wasserman, L.H., Burgett, W.S., Chambers, K.C., Heasley, J.N., Kaiser, N., Magnier, E.A., Morgan, J.S., Price, P.A., Jørgensen, U.G., Dominik, M., Hinse, T., Sahu, K., Snodgrass, C., 2012. Observational and dynamical characterization of main-belt comet P/2010 R2 (La Sagra). Astron. J. 143 (5), 104. http://dx.doi.org/10.1088/0004-6256/143/5/104, arXiv:1109.6350.

Hui, M.-T., Jewitt, D., Clark, D., 2018. Pre-discovery observations and orbit of comet C/2017 K2 (PANSTARRS). Astron. J. 155 (1), 25. http://dx.doi.org/10.3847/1538-3881/aa9be1, arXiv:1711.06355.

Ivanova, A.V., Afanasiev, V.L., Korsun, P.P., Baranskii, A.R., Andreev, M.V., Ponomarenko, V.A., 2012. The rotation period of comet 29P/Schwassmann–Wachmann 1 determined from the dust structures (Jets) in the coma. Sol. Syst. Res. 46 (4), 313–319. http://dx.doi.org/10.1134/S003809461204003X, arXiv:2012.09007.

Ivanova, A.V., Korsun, P.P., Afanasiev, V.L., 2009. Photometric investigations of distant comets C/2002 VQ94 (Linear) and 29P/Schwassmann–Wachmann-1. Sol. Syst. Res. 43 (5), 453–462. http://dx.doi.org/10.1134/S0038094609050086, arXiv:2012.09008.

Ivanova, O.V., Luk'yanyk, I.V., Kiselev, N.N., Afanasiev, V.L., Picazzio, E., Cavichia, O., de Almeida, A.A., Andrievsky, S.M., 2016. Photometric and spectroscopic analysis of comet 29P/Schwassmann–Wachmann 1 activity. Planet. Space Sci. 121, 10–17. http://dx.doi.org/10.1016/j.pss.2015.12.001, arXiv:2012.10705.

Ivanova, O.V., Picazzio, E., Luk'yanyk, I.V., Cavichia, O., Andrievsky, S.M., 2018. Spectroscopic observations of the comet 29P/Schwassmann–Wachmann 1 at the SOAR telescope. Planet. Space Sci. 157, 34–38. http://dx.doi.org/10.1016/j.pss.2018.04.003, arXiv:1805.06715.

Ivanova, O., Rosenbush, V., Luk'yanyk, I., Kolokolova, L., Kleshchonok, V., Kiselev, N., Afanasiev, V., Renée Kirk, Z., 2021. Observations of distant comet C/2011 KP36 (spacewatch): Photometry, spectroscopy, and polarimetry. Astron. Astrophys. 651, A29. http://dx.doi.org/10.1051/0004-6361/202039668.

Ivanova, O.V., Skorov, Y.V., Korsun, P.P., Afanasiev, V.L., Blum, J., 2011. Observations of the long-lasting activity of the distant comets 29P Schwassmann–Wachmann 1, C/2003 WT42 (Linear) and C/2002 VQ94 (Linear). Icarus 211 (1), 559–567. http://dx.doi.org/10.1016/j.icarus.2010.10.026, arXiv:2102.04829.

Jewitt, D., 1990. The persistent coma of comet P/Schwassmann–Wachmann 1. Astrophys. J. 351, 277. http://dx.doi.org/10.1086/168463.

Jewitt, D., 1991. Cometary photometry. In: Newburn, J., Neugebauer, M., Rahe, J. (Eds.), IAU Colloq. 116: Comets in the Post-Halley Era. In: Astrophysics and Space Science Library, vol. 167, p. 19. http://dx.doi.org/10.1007/978-94-011-3378-4_2.

Jewitt, D., 2005. A first look at the damocloids. Astron. J. 129 (1), 530–538. http://dx.doi.org/10.1086/426328.

Jewitt, D., 2009. The active centaurs. Astron. J. 137 (5), 4296–4312. http://dx.doi.org/10.1088/0004-6256/137/5/4296, arXiv:0902.4687.

Jewitt, D., Kim, Y., Luu, J., Graykowski, A., 2019. The discus comet: C/2014 B1 (Schwartz). Astron. J. 157 (3), 103. http://dx.doi.org/10.3847/1538-3881/aafe05, arXiv:1901.01438.

Keane, J., DiSanti, M.D., Bonev, B.P., Paganini, L., Meech, K.J., Mumma, M.J., Villanueva, G.L., 2013. CO detections in 29P/Schwassmann–Wachmann 1 AT 6.21 AU using NIRSPEC. In: AAS/Division for Planetary Sciences Meeting Abstracts #45. In: AAS/Division for Planetary Sciences Meeting Abstracts, vol. 45, p. 413.23.

Kokhirova, G.I., Ivanova, O.V., Rakhmatullaeva, F.D., Buriev, A.M., Khamroev, U.K., 2020. Astrometric and photometric observations of comet 29P/Schwassmann–Wachmann 1 at the Sanglokh international astronomical observatory. Planet. Space Sci. 181, 104794. http://dx.doi.org/10.1016/j.pss.2019.104794, arXiv:2012.06833.

Kokhirova, G.I., Ivanova, O.V., Rakhmatullaeva, F.D., Khamroev, U.K., Buriev, A.M., Abdulloev, S.K., 2018. Results of complex observations of asteroid (596) scheila at the Sanglokh international astronomical observatory. Sol. Syst. Res. 52 (6), 495–504. http://dx.doi.org/10.1134/S0038094618060035, arXiv:2012.08430.

Korsun, P.P., Ivanova, O.V., Afanasiev, V.L., Kulyk, I.V., 2016. Distant Jupiter family comet P/2011 P1 (McNaught). Icarus 266, 88–95.

Korsun, P.P., Kulyk, I., Ivanova, O.V., Zakhozhay, O.V., Afanasiev, V.L., Sergeev, A.V., Velichko, S.F., 2016. Optical spectrophotometric monitoring of comet C/2006 W3 (Christensen) before perihelion. Astron. Astrophys. 596, A48. http://dx.doi.org/10.1051/0004-6361/201629046.

Korsun, P.P., Rousselot, P., Kulyk, I.V., Afanasiev, V.L., Ivanova, O.V., 2014. Distant activity of comet C/2002 VQ94 (Linear): Optical spectrophotometric monitoring between 8.4 and 16.8 au from the sun. Icarus 232, 88–96. http://dx.doi.org/10.1016/j.icarus.2014.01.006, arXiv:1401.3137.

Lamy, P.L., Toth, I., Weaver, H.A., A'Hearn, M.F., Jorda, L., 2009. Properties of the nuclei and comae of 13 ecliptic comets from hubble space telescope snapshot observations. Astron. Astrophys. 508 (2), 1045–1056. http://dx.doi.org/10.1051/0004-6361/200811462.

Larson, S.M., Sekanina, Z., 1984. Coma morphology and dust-emission pattern of periodic comet halley. I - high-resolution images taken at Mount Wilson in 1910. Astron. J. 89, 571–578. http://dx.doi.org/10.1086/113551.

Lowry, S.C., Fitzsimmons, A., 2001. CCD photometry of distant comets II. Astron. Astrophys. 365, 204–213. http://dx.doi.org/10.1051/0004-6361:20000180.

Lowry, S.C., Fitzsimmons, A., 2005. William herschel telescope observations of distant comets. Mon. Not. RAS 358 (2), 641–650. http://dx.doi.org/10.1111/j.1365-2966.2005.08825.x.

Lowry, S.C., Fitzsimmons, A., Cartwright, I.M., Williams, I.P., 1999. CCD photometry of distant comets. Astron. Astrophys. 349, 649–659.

Lowry, S.C., Fitzsimmons, A., Collander-Brown, S., 2003. CCD photometry of distant comets. III. Ensemble properties of Jupiter-family comets. Astron. Astrophys. 397, 329–343. http://dx.doi.org/10.1051/0004-6361:20021486.

Lowry, S.C., Weissman, P.R., 2003. CCD observations of distant comets from Palomar and Steward observatories. Icarus 164 (2), 492–503. http://dx.doi.org/10.1016/S0019-1035(03)00129-5.

Luu, J., Jewitt, D., 1993. Periodic comet Schwassmann–Wachmann 1. IAU Cirulars 5692, 3.

Martino, S., Tancredi, G., Monteiro, F., Lazzaro, D., Rodrigues, T., 2019. Monitoring of asteroids in cometary orbits and active asteroids. Planet. Space Sci. 166, 135–148. http://dx.doi.org/10.1016/j.pss.2018.09.001.

Mazzotta Epifani, E., Dall'Ora, M., di Fabrizio, L., Licandro, J., Palumbo, P., Colangeli, L., 2010. The activity of comet C/2007 D1 (Linear) at 9.7 AU from the sun. Astron. Astrophys. 513, A33. http://dx.doi.org/10.1051/0004-6361/200913535.

Mazzotta Epifani, E., Palumbo, P., Capria, M.T., Cremonese, G., Fulle, M., Colangeli, L., 2007. The distant activity of short-period comets - I. Mon. Not. RAS 381 (2), 713–722. http://dx.doi.org/10.1111/j.1365-2966.2007.12181.x.

Mazzotta Epifani, E., Palumbo, P., Capria, M.T., Cremonese, G., Fulle, M., Colangeli, L., 2009. The distant activity of the long period comets C/2003 O1 (Linear) and C/2004 K1 (Catalina). Astron. Astrophys. 502 (1), 355–365. http://dx.doi.org/10.1051/0004-6361/200811527.

Meech, K.J., Belton, M.J.S., Mueller, B.E.A., Dicksion, M.W., Li, H.R., 1993. Nucleus properties of P/Schwassmann–Wachmann 1. Astron. J. 106, 1222. http://dx.doi.org/10.1086/116721.

Meech, K.J., Pittichová, J., Bar-Nun, A., Notesco, G., Laufer, D., Hainaut, O.R., Lowry, S.C., Yeomans, D.K., Pitts, M., 2009. Activity of comets at large heliocentric distances pre-perihelion. Icarus 201 (2), 719–739. http://dx.doi.org/10.1016/j.icarus.2008.12.045.

Miles, R., 2016. Discrete sources of cryovolcanism on the nucleus of comet 29P/Schwassmann–Wachmann and their origin. Icarus 272, 387–413. http://dx.doi.org/10.1016/j.icarus.2015.11.011.

Miles, R., Faillace, G.A., Mottola, S., Raab, H., Roche, P., Soulier, J.-F., Watkins, A., 2016. Anatomy of outbursts and quiescent activity of comet 29P/Schwassmann–Wachmann. Icarus 272, 327–355. http://dx.doi.org/10.1016/j.icarus.2015.11.019.

Neslusan, L., Ivanova, O., Husarik, M., Svoren, J., Krisandova, Z.S., 2016. Dust productivity and impact collision of the asteroid (596) Scheila. Planet. Space Sci. 125, 37–42. http://dx.doi.org/10.1016/j.pss.2016.01.017, arXiv:2012.08434.

Picazzio, E., Luk'yanyk, I.V., Ivanova, O.V., Zubko, E., Cavichia, O., Videen, G., Andrievsky, S.M., 2019. Comet 29P/Schwassmann–Wachmann 1 dust environment from photometric observation at the SOAR telescope. Icarus 319, 58–67. http://dx.doi.org/10.1016/j.icarus.2018.09.028, arXiv:2012.10718.

Rosenbush, V., Ivanova, O., Kleshchonok, V., Kiselev, N., Afanasiev, V., Shubina, O., Petrov, D., 2020. Comet 2P/Encke in apparitions of 2013 and 2017: I. Imaging photometry and long-slit spectroscopy. Icarus 348, 113767. http://dx.doi.org/10.1016/j.icarus.2020.113767, arXiv:2101.04172.

Rosenbush, V., Kiselev, N., Husárik, M., Ivanova, O., Luk'yanyk, I., Kleshchonok, V., Tomko, D., Kaňuchová, Z., Pit, N., Antoniuk, K., Karpov, N., Savushkin, A., Zhuzhulina, E., 2021. Photometry and polarimetry of comet 46P/Wirtanen in the 2018 apparition. Mon. Not. RAS 503 (3), 4297–4308. http://dx.doi.org/10.1093/mnras/stab585.

Rousselot, P., Korsun, P.P., Kulyk, I., Guilbert-Lepoutre, A., Petit, J.-M., 2016. A long-term follow up of 174P/Echeclus. Mon. Not. RAS 462, S432–S442. http://dx.doi.org/10.1093/mnras/stw3054.

Rousselot, P., Kryszczyńska, A., Bartczak, P., Kulyk, I., Kamiński, K., Dudziński, G., Anderson, S.E., Noyelles, B., Guilbert-Lepoutre, A., 2021. New constraints on the physical properties and dynamical history of centaur 174P/Echeclus. Mon. Not. RAS 507 (3), 3444–3460. http://dx.doi.org/10.1093/mnras/stab2379.

Samarasinha, N.H., Larson, S.M., 2014. Image enhancement techniques for quantitative investigations of morphological features in cometary comae: A comparative study. Icarus 239, 168–185. http://dx.doi.org/10.1016/j.icarus.2014.05.028, arXiv:1406.0033.

Schambeau, C.A., Fernández, Y.R., Samarasinha, N.H., Mueller, B.E.A., Woodney, L.M., 2017. Analysis of R-band observations of an outburst of comet 29P/Schwassmann–Wachmann 1 to place constraints on the nucleus' rotation state. Icarus 284, 359–371. http://dx.doi.org/10.1016/j.icarus.2016.11.026.

Schleicher, D.G., Lederer, S.M., Millis, R.L., Farnham, T.L., 1997. Photometric behavior of Comet Hale-Bopp (C/1995 O1) before perihelion. Science 275, 1913–1915. http://dx.doi.org/10.1126/science.275.5308.1913.

Schleicher, D.G., Millis, R.L., Birch, P.V., 1998. Narrowband photometry of comet P/Halley: Variation with heliocentric distance, season, and solar phase angle. Icarus 132 (2), 397–417. http://dx.doi.org/10.1006/icar.1997.5902.

Sekanina, Z., 1988a. Outgassing asymmetry of periodic comet encke. II. Apparitions 1868–1918 and a study of the nucleus evolution. Astron. J. 96, 1455. http://dx.doi.org/10.1086/114897.

Sekanina, Z., 1988b. Outgassing symmetry of periodic comet encke. I. Apparitions 1924–1984. Astron. J. 95, 911. http://dx.doi.org/10.1086/114689.

Shi, J., Ma, Y., Liang, H., Xu, R., 2019. Research of activity of main belt comets 176P/Linear, 238P/read and 288P/(300163) 2006 VW$_{139}$. Sci. Rep. 9, 5492. http://dx.doi.org/10.1038/s41598-019-41880-0.

Shi, J.C., Ma, Y.H., Zheng, J.Q., 2014. CCD photometry of distant active comets 228P/LINEAR, C/2006 S3 (LONEOS) and 29P/Schwassmann–Wachmann 1. Mon. Not. RAS 441 (1), 739–744. http://dx.doi.org/10.1093/mnras/stu607.

Sierks, H., Barbieri, C., Lamy, P.L., Rodrigo, R., Koschny, D., Rickman, H., Keller, H.U., Agarwal, J., A'Hearn, M.F., Angrilli, F., Auger, A.-T., Barucci, M.A., Bertaux, J.-L., Bertini, I., Besse, S., Bodewits, D., Capanna, C., Cremonese, G., Da Deppo, V., Davidsson, B., Debei, S., De Cecco, M., Ferri, F., Fornasier, S., Fulle, M., Gaskell, R., Giacomini, L., Groussin, O., Gutierrez-Marques, P., Gutiérrez, P.J., Güttler, C., Hoekzema, N., Hviid, S.F., Ip, W.-H., Jorda, L., Knollenberg, J., Kovacs, G., Kramm, J.R., Kührt, E., Küppers, M., La Forgia, F., Lara, L.M., Lazzarin, M., Leyrat, C., Lopez Moreno, J.J., Magrin, S., Marchi, S., Marzari, F., Massironi, M., Michalik, H., Moissl, R., Mottola, S., Naletto, G., Oklay, N., Pajola, M., Pertile, M., Preusker, F., Sabau, L., Scholten, F., Snodgrass, C., Thomas, N., Tubiana, C., Vincent, J.-B., Wenzel, K.-P., Zaccariotto, M., Pätzold, M., 2015. On the nucleus structure and activity of comet 67P/Churyumov–Gerasimenko. Science 347 (6220), aaa1044. http://dx.doi.org/10.1126/science.aaa1044.

Snodgrass, C., Lowry, S.C., Fitzsimmons, A., 2006. Photometry of cometary nuclei: Rotation rates, colours and a comparison with kuiper belt objects. Mon. Not. RAS 373 (4), 1590–1602. http://dx.doi.org/10.1111/j.1365-2966.2006.11121.x, arXiv:astro-ph/0610301.

Snodgrass, C., Lowry, S.C., Fitzsimmons, A., 2008. Optical observations of 23 distant Jupiter family comets, including 36P/Whipple at multiple phase angles. Mon. Not. RAS 385 (2), 737–756. http://dx.doi.org/10.1111/j.1365-2966.2008.12900.x, arXiv:0712.4204.

Stansberry, J.A., Van Cleve, J., Reach, W.T., Cruikshank, D.P., Emery, J.P., Fernandez, Y.R., Meadows, V.S., Su, K.Y.L., Misselt, K., Rieke, G.H., Young, E.T., Werner, M.W., Engelbracht, C.W., Gordon, K.D., Hines, D.C., Kelly, D.M., Morrison, J.E., Muzerolle, J., 2004. Spitzer observations of the dust coma and nucleus of 29P/Schwassmann–Wachmann 1. Astrophys. J. Suppl. 154 (1), 463–468. http://dx.doi.org/10.1086/422473.

Szabó, G.M., Csák, B., Sárneczky, K., Kiss, L.L., 2001. Photometric observations of distant active comets. Astron. Astrophys. 374, 712–718. http://dx.doi.org/10.1051/0004-6361:20010715, arXiv:astro-ph/0105336.

Trigo-Rodríguez, J.M., García-Hernández, D.A., Rodríguez, D., Sánchez, A., Lacruz, J., 2012. Centaur 29P/Schwassmann–Wachmann 1: Photometric activity during 2012. In: European Planetary Science Congress 2012. Copernicus.org, pp. EPSC2012–493.

Trigo-Rodríguez, J.M., García-Hernández, D.A., Sánchez, A., Lacruz, J., Davidsson, B.J.R., Rodríguez, D., Pastor, S., de Los Reyes, J.A., 2010. Outburst activity in comets - II. A multiband photometric monitoring of comet 29P/Schwassmann–Wachmann 1. Mon. Not. RAS 409 (4), 1682–1690. http://dx.doi.org/10.1111/j.1365-2966.2010.17425.x, arXiv:1009.2381.

Voitko, A., Zubko, E., Ivanova, O., Luk'yanyk, I., Kochergin, A., Husárik, M., Videen, G., 2022. Color variations of comet 29P/Schwassmann–Wachmann 1 in 2018. Icarus 115236. http://dx.doi.org/10.1016/j.icarus.2022.115236, URL https://www.sciencedirect.com/science/article/pii/S0019103522003293.

Whipple, F.L., 1980. Rotation and outbursts of comet P/Schwassmann–Wachmann 1. Astron. J. 85, 305–313. http://dx.doi.org/10.1086/112676.

Wong, I., Mishra, A., Brown, M.E., 2019. Photometry of active centaurs: Colors of dormant active centaur nuclei. Astron. J. 157 (6), 225. http://dx.doi.org/10.3847/1538-3881/ab1b22, arXiv:1904.09255.

Zacharias, N., Finch, C.T., Girard, T.M., Henden, A., Bartlett, J.L., Monet, D.G., Zacharias, M.I., 2013. The fourth US naval observatory CCD astrograph catalog (UCAC4). Astron. J. 145 (2), 44. http://dx.doi.org/10.1088/0004-6256/145/2/44, arXiv:1212.6182.